\documentclass[a4paper]{mem} \usepackage{natbib} \usepackage{graphicx}

\usepackage{indentfirst}

\usepackage{amsmath}

\newcommand{\de}{\partial}
\newcommand{\be}{\begin{equation}}
\newcommand{\ee}{\end{equation}}
\newcommand{\bea}{\begin{eqnarray}}
\newcommand{\eea}{\end{eqnarray}}
\begin{document}
\setcounter{page}{210}

\title{A new three-dimensional general-relativistic hydrodynamics code
}

\author{L. Baiotti \inst{1,2}, I. Hawke \inst{3}, P.J. Montero \inst{1}, L. Rezzolla \inst{1,2} 
}

\institute{SISSA, Via Beirut 2-4, 34014 Trieste, Italy\\
\and  INFN, Sezione di Trieste, Via Beirut 2-4, 34014 Trieste, Italy\\
\and Albert-Einstein-Institut, Am M\"{u}hlenberg 1, 14476 Golm, Germany }

\abstract{ 
We present a new three-dimensional general relativistic hydrodynamics
code, the {\tt Whisky} code. This code incorporates the expertise
developed over the past years in the numerical solution of Einstein
equations and of the hydrodynamics equations in a curved spacetime,
and is the result of a collaboration of several European
Institutes. We here discuss the ability of the code to carry out
long-term accurate evolutions of the linear and nonlinear dynamics of
isolated relativistic stars.


   }
\authorrunning{L. Baiotti, I. Hawke, P. Montero and L. Rezzolla}
\titlerunning{The {\tt Whisky} code}
   \maketitle
%

\section{Introduction}

During the last few years, computational general relativistic
astrophysics has become increasingly important and accurate. This is
partly due to the rapid increase in computing power through massively
parallel supercomputers which make large-scale, multi-dimensional
numerical simulations possible~\citep{papI, shibata00, papII, shibata02,
duez}. In addition to being a unique tool for investigating General
Relativity in regimes of strong and rapidly varying gravitational fields,
such simulations are also needed to fully understand the incoming wealth
of observations from high-energy astronomy and (near-future)
gravitational wave astronomy. In an attempt to respond, at least in part,
to this need we have recently developed {\tt Whisky}, a three-dimensional
(3D) general relativistic hydrodynamics code. The {\tt Whisky} code is
the result of an ongoing collaboration among several European Institutes,
i.e. the Albert Einstein Institute (Golm, Germany), SISSA (Trieste,
Italy) and the Universities of Thessaloniki (Greece) and Valencia
(Spain), joined in a European Network investigating sources of
gravitational waves (see {\tt www.eu-network.org} for further
information).

In practice, the {\tt Whisky} code solves the general relativistic
hydrodynamics equations on a 3D numerical grid with Cartesian
coordinates. The code has been constructed within the framework of the
{\tt Cactus Computational Toolkit} (see {\tt www.cactuscode.org} for
details), developed at the Albert Einstein Institute (Golm), which
provides high-level facilities such as parallelization, input/output,
etc., but also the solution of the Einstein equations with matter
terms being provided by the {\tt Whisky} code.

In this paper we briefly discuss the main features of {\tt Whisky} and
present some tests of its validation. Indeed, validation represents an
important aspect of the development of any modern 3D finite-difference
code. The reasons for this are rather simple and are related to: {\it
(i)} the lack of precise knowledge of the space of solutions of the
coupled system of the Einstein and general relativistic hydrodynamics
equations; {\it (ii)} the likely chance that coding errors are made in
the implementation of the thousands of terms involved in the solution
of such a complicated set of coupled partial differential equations;
{\it (iii)} the complexity of the computational infrastructure needed
for the use of the code in a massively parallel environment which
increases the risk of computational errors.

The tests presented here will show both the accuracy and the
convergence of our formulation for the general relativistic
hydrodynamics equations, which are coupled to a conformal
transverse-traceless formulation of the Einstein equations
\citep{nakamura}. They will also show the ability of the code
to follow stably the linear and nonlinear dynamics of isolated
relativistic stars. More specifically, we will first present results
of the linear pulsations of spherical and rapidly rotating stars. The
computed frequencies of radial and quasi-radial oscillations will be
compared with the corresponding frequencies obtained with
lower-dimensional numerical codes, or with alternative techniques such
as the Cowling approximation (in which the spacetime is held fixed and
only the general relativistic hydrodynamics equations are evolved), or
with relativistic perturbative methods. The comparison shows an
excellent agreement confirming the ability of the code to extract
physically relevant information even from tiny perturbations. The
successful determination of the eigenfrequencies for rapidly rotating
stars computed with our code is particularly noteworthy since the
equivalent problem has not yet been tackled with perturbative
techniques.

We will also investigate the nonlinear dynamics of stellar models that
are unstable to the fundamental radial mode of pulsation. We show that
upon perturbation, the unstable models will either collapse to a black
hole, or migrate to a configuration in the stable branch of
equilibrium configurations. In the case of a gravitational collapse,
we will follow the evolution all the way down to the formation of a
black hole, tracking the generation of its apparent horizon. In the
case of migration to the stable branch, on the other hand, we will be
able to accurately follow the nonlinear oscillations that accompany
this process and that can give rise to strong shocks. The ability to
simulate large amplitude oscillations is important as we expect a
neutron star formed in a supernova core-collapse \citep{SN1, SN2} or
in the accretion-induced collapse of a white dwarf to oscillate
violently in its early stages of life.

We use the signature \mbox{$(-, +, +, +)$} and units in which \mbox{$c
= G = M_\odot  = 1$}. Greek indices are taken to run from 0 to 3 and
Latin indices from 1 to 3.

\section{Basic Equations}

The {\tt Whisky} code has been constructed exploiting the expertise
developed in the building of a similar but distinct code which has been
described extensively in \citep{papI, papII}. The development of the {\tt
Whisky} code has benefited from the release of a public version of the
general relativistic hydrodynamics code described in Font et al. (2000b),
(2002), and developed mostly by the group at the Washington University as
part of the NASA Neutron Star Grand Challenge Project (see {\tt
wugrav.wustl.edu/Codes/GR3D} for details).

	The {\tt Whisky} code, however, also incorporates important
recent developments regarding, in particular, new numerical methods for
the solution of the hydrodynamics equations. These include: {\it (i)} the
Piecewise Parabolic Method (PPM) \citep{PPM} and the Essentially
Non-Oscillatory (ENO) methods \citep{ENO} for the cell reconstruction
procedure; {\it (ii)} the Harten-Lax-van Leer-Einfeldt (HLLE)
\citep{HLLE} approximate Riemann solver, the Marquina flux formula
\citep{marquina}; {\it (iii)} the analytic expression for the left
eigenvectors \citep{anal left} and compact flux formulae \citep{compact
flux} for the Roe Riemann solver and the one using the Marquina flux
formula; {\it (iv)} the possibility to couple the general relativistic
hydrodynamics equations with a conformally decomposed 3-metric. The
incorporation of these new numerical techniques in the code has lead to a
much improved ability to simulate relativistic stars, as will be shown in
the section devoted to the tests.  The interested reader may also refer
to \cite{papI, papII} for more details about the general formulation and
the results obtained with the similar, distinct code.

While the {\tt Cactus} code provides at each time step a solution of
the Einstein equations (see \cite{alcubierre et al} for the validation
of the {\tt Cactus} code for the spacetime evolution) 
\be
G_{\mu \nu}=8\pi T_{\mu \nu}\ , 
\ee
where $G_{\mu \nu}$ is the Einstein tensor and $T_{\mu \nu}$ is the
stress-energy tensor, the {\tt Whisky} code provides the time evolution
of the hydrodynamics equations, expressed through the conservation
equations for the stress-energy tensor $T^{\mu\nu}$ and for the matter
current density $J^\mu$
\be
\label{hydro eqs}
\nabla_\mu T^{\mu\nu} = 0\;,\;\;\;\;\;\;
\nabla_\mu J^\mu = 0.
\ee

An important feature of the {\tt Whisky} code is the implementation of
the so called \textit{Valencia formulation} of the hydrodynamics
equations \citep{marti ibanez miralles 91, banyuls, anal left}, in which the set of
equations (\ref{hydro eqs}) is written in hyperbolic,
flux-conservative form
\begin{equation}
  \label{eq:consform1}
  \partial_t {\mathbf q} + \de_i {\mathbf f}^{(i)} ({\mathbf q}) = {\mathbf
    s} ({\mathbf q}),
\end{equation}
where the right hand side (the source terms) depends only on the
metric and on the stress-energy tensor, vanishing in a flat spacetime,
where the strict hyperbolicity is recovered. In order to write system
(\ref{hydro eqs}) in the form of system (\ref{eq:consform1}), the
\textit{primitive} hydrodynamical variables (i.e. the rest
mass-density $\rho$, the pressure $p$, the fluid 3-velocities $v^i$,
the internal energy density $\epsilon$ and the Lorentz factor $W$) are
mapped to the so called \textit{conserved} variables
\mbox{${\mathbf q} \equiv (D, S^i, \tau)$} via the relations

\vspace{-0.2 cm}
\begin{eqnarray}
  \label{eq:prim2con}
   D &\equiv& \sqrt{\gamma}W\rho\ , \nonumber \\
   S^i &\equiv& \sqrt{\gamma} \rho h W^2 v^i\ , \nonumber \\
   \tau &\equiv& \sqrt{\gamma}\left( \rho h W^2 - p\right) - D\ , 
\end{eqnarray}
where $\gamma$ is the determinant of the spatial 3-metric
$\gamma_{ij}$ and $h \equiv 1 + \epsilon + p/\rho$ is the specific
enthalpy.  The Lorentz factor is defined in terms of the velocities
and of the 3-metric as $W = (1-\gamma_{ij}v^i v^j)^{-1/2}$. Note that
only five of the primitive variables are independent.

	Finally, an equation of state is used to relate the pressure
to the rest mass density and to the energy density. The code can use
any equation of state, but presently only a polytropic equation of state of the type

\be
\label{poly}
p = K \rho^{\Gamma}\ ,
\ee
an ``ideal fluid'' equation of state
\be
\label{id fluid}
p = (\Gamma-1) \rho\, \epsilon
\ee
and a hybrid equation of state as described in \cite{hybrid eos} have
been implemented. For more details on this formulation, see also the
review by \cite{livrevFont}.

An important feature of the Valencia formulation is that it is
possible to extend to relativistic hydrodynamics the powerful
numerical methods developed in classical hydrodynamics; in particular,
our code takes advantage of the High Resolution Shock Capturing (HRSC)
properties of Godunov type methods \citep{godunov}.  For a full
introduction to HRSC methods see~\cite{laney}, \cite{toro} and
\cite{leveque}.

\section{Numerical Techniques}

The time update of all the equations, general relativistic
hydrodynamics and Einstein, are performed with the {\it Method of
Lines} (MoL) \citep{laney,toro}. The method of lines is a procedure to
transform a set of  partial differential equations such as
(\ref{eq:consform1}) into a set of  ordinary
differential equations. This is done by integrating equations (\ref{eq:consform1}) over space in every
computational cell defined by its position $(x_i, y_j, z_k )$
\begin{eqnarray}
  \label{eq:molrhs1} \nonumber
  &&\frac{{\rm d} }{{\rm d} t} (\widetilde{{\mathbf q}}) = {\mathbf L}(\widetilde{{\mathbf q}})  =  \int \!\!\!\! \int \!\!\!\!
      \int {\mathbf s} \,{\rm d}^3 x + \\
&+&\!\! \int_{y_{j-1/2}}^{y_{j+1/2}}\!\!\!\!
    \int_{z_{k-1/2}}^{z_{k+1/2}}\!\!\!\! {\mathbf f}^{(1)}({\mathbf q}
    (x_{i-1/2}, y, z)) {\rm d} y \, {\rm d} z \nonumber\\
   &-&\!\! \int_{y_{j-1/2}}^{y_{j+1/2}}\!\!\!\! 
    \int_{z_{k-1/2}}^{z_{k+1/2}} \!\!\!\! {\mathbf f}^{(1)} ({\mathbf q}
    (x_{i+1/2}, y, z)) {\rm d} y \, {\rm d} z \nonumber\\
 &+& \ldots \ ,
\end{eqnarray}
where $\widetilde{{\mathbf q}}$ is the spatially integrated vector of conserved
variables, i.e. 
\be
\widetilde{{\mathbf q}} \equiv \int  {\mathbf q}\ \  {\rm d} x\, {\rm d} y\, {\rm d} z 
\ee
and ${\mathbf f}^{(i)}$ is the $i$-th component of the flux five-vector ${\mathbf f}$.

Several time-integrators are available in our implementation
of MoL and the order of
accuracy of the solution of the ordinary differential
equation~(\ref{eq:molrhs1}) is the same as the truncation order of the
integrator employed, provided that the discrete operator ${\mathbf L}$ is
of the same order in space and at least first-order accurate in time.

In our implementation of MoL, the right hand side operator ${\mathbf
L}(\widetilde{\mathbf q})$ is simplified by approximating the integrals
with the midpoint rule to get
\begin{equation}
  \label{eq:molrhs2}
  {\mathbf L}(\widetilde{{\mathbf q}}) = {\mathbf s}_{i,j,k} + {\mathbf f}^{(1)}_{i-1/2,j,k} -
    {\mathbf f}^{(1)}_{i+1/2,j,k} + \ldots\ \ \ .
\end{equation}
 Given this simplification, the calculation
of the right hand side of (\ref{eq:molrhs1}) splits into the following parts:
\begin{enumerate}

\item Calculation of the {\it source terms} ${\mathbf
s}({\mathbf q}(x_i, y_j, z_k))$ at all the grid points.
 
\item {\it Reconstruction} of the data ${\mathbf q}$ to both sides of
a cell boundary. In this way, two values ${\mathbf q}_{_L}$ and
${\mathbf q}_{_R}$ of ${\mathbf q}_{i+1/2,j,k}$ are determined at cell
boundary; ${\mathbf q}_{_L}$ is obtained from cell $i$ (left cell) and
${\mathbf q}_{_R}$ from cell $i+1$ (right cell).  The code implements
several reconstruction methods. In particular, as Total Variation
Diminishing (TVD) methods we have implemented ``minmod'', van Leer
monotonized centered \citep{vanLeer} and Superbee
\citep{toro}. Additional reconstruction methods are: arbitrary order
ENO methods \citep{ENO} and the Piecewise Parabolic Method
\citep{PPM}, which is a third order accurate in space. As mentioned
below, PPM has emerged as our actual best choice for all the test
evolutions we present here.

\item Solution at cell boundary of the {\it Riemann problem}
\citep{leveque,toro,laney} having the values ${\mathbf q}_{_{L,R}}$ as
initial data.
 
\item Calculation in each coordinate direction ($x, y, z$) of the {\it
inter-cell flux} ${\mathbf f}^{(x)}({\mathbf q}_{i+1/2,j,k})$, ${\mathbf f}^{(y)}({\mathbf q}_{i,j+1/2,k})$,
${\mathbf f}^{(z)}({\mathbf q}_{i,j,k+1/2})$, that is the
flux across the boundary between a cell (e.g. the $i$-th) and its
closest neighbour (e.g. the $(i+1)$-th).

\item {\it Recovery} of the primitive variables and computation of the
stress-energy tensor for use in the Einstein equations.

\end{enumerate}

As a result of steps 1. -- 4., the core of the {\tt Whisky} code is
effectively represented by two routines. One that reconstructs the
function ${\mathbf q}$ at the boundaries of a
computational cell and another one that calculates the inter-cell flux
${\mathbf f}$ at this cell boundary.

As for the reconstruction methods, a similar variety is present for the
approximate Riemann solvers implemented in the {\tt Whisky} code. More
specifically, are available the fast HLLE \citep{HLLE} and the widespread
Roe \citep{roe solver} solvers as well as the accurate Marquina flux
formula \citep{marquina}, which is used to solve the Riemann problem in a
way that differs from the Roe solver only at sonic points, where the Roe
solver has problems. The Roe-based approximate Riemann solvers need the
computation of the eigenvalues and eigenvectors (from both the right and
left cell) of the linearized Jacobian matrices ${\mathbf A}_{_L}$ and
${\mathbf A}_{_R}$ given by \mbox{${\mathbf f}_{_L} = {\mathbf
A}_{_L}{\mathbf q}_{_L}$} and \mbox{${\mathbf f}_{_R} = {\mathbf
A}_{_R}{\mathbf q}_{_R}$}. We have implemented the analytic expression
for the left eigenvectors \citep{anal left}, thus avoiding the
computationally expensive inversion of the three $5\times 5$ matrices of
the right eigenvectors, associated to each spatial direction. This
implementation brings a 40\% reduction of the computational time spent in
the solution of the hydrodynamics equations. However, in evolutions
involving also the time integration of the Einstein equations, this is
reduced to a 5\% decrease in computational cost. This is due to the fact
that the largest part of the time is spent in the update of the spacetime
field variables.

   \begin{figure}
   \centering
\includegraphics[angle=-90,width=6cm]{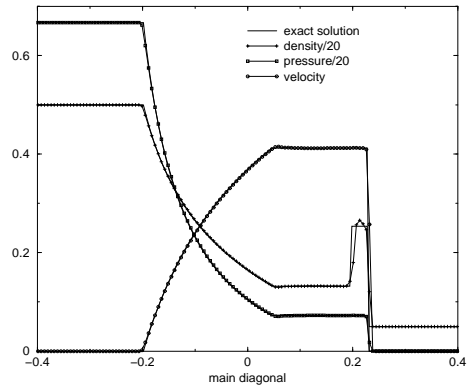} \caption{Solution of
     a Riemann problem set on the main diagonal of the cubic grid. The
     figure shows the comparison of the hydrodynamical variables evolved
     by {\tt Whisky}, indicated with symbols, with the exact
     solution. The numerical simulation was obtained with the van Leer
     reconstruction method and the Roe solver, on a $140^3$ grid.}
     \label{figShock} \end{figure}

\section{Numerical Tests}

As mentioned above several tests have been performed to assess the
stability and accuracy of our code. The results obtained so far are very
encouraging and already during these preliminary steps some new physical
results have been achieved.

First of all, we consider a standard shock tube test, setting as initial
data a global Riemann problem, i.e. one in which the initial
discontinuity is orthogonal to the main diagonal of the cubic grid. More
precisely the initial data consist of a left and right state with values
\begin{alignat}{5}
\rho_R &=& 1;\;\; p_R &= 1.666 \times 10^{-6};\;\; &v_R &= 0\nonumber\\
\rho_L &=& 10;\;\; p_L &= 1.333; \;\;&v_L &= 0\nonumber
\end{alignat}

In Fig.~\ref{figShock} we show the solution at a given time together
with the exact solution. The excellent agreement of the two sets of
curves is particularly remarkable if one bears in mind that the
initial shock is placed on the main diagonal of the cubic grid, so
that the evolution is fully 3-dimensional.

\begin{figure*}
   \centering
   \resizebox{\hsize}{!}{\rotatebox[]{0}{\includegraphics{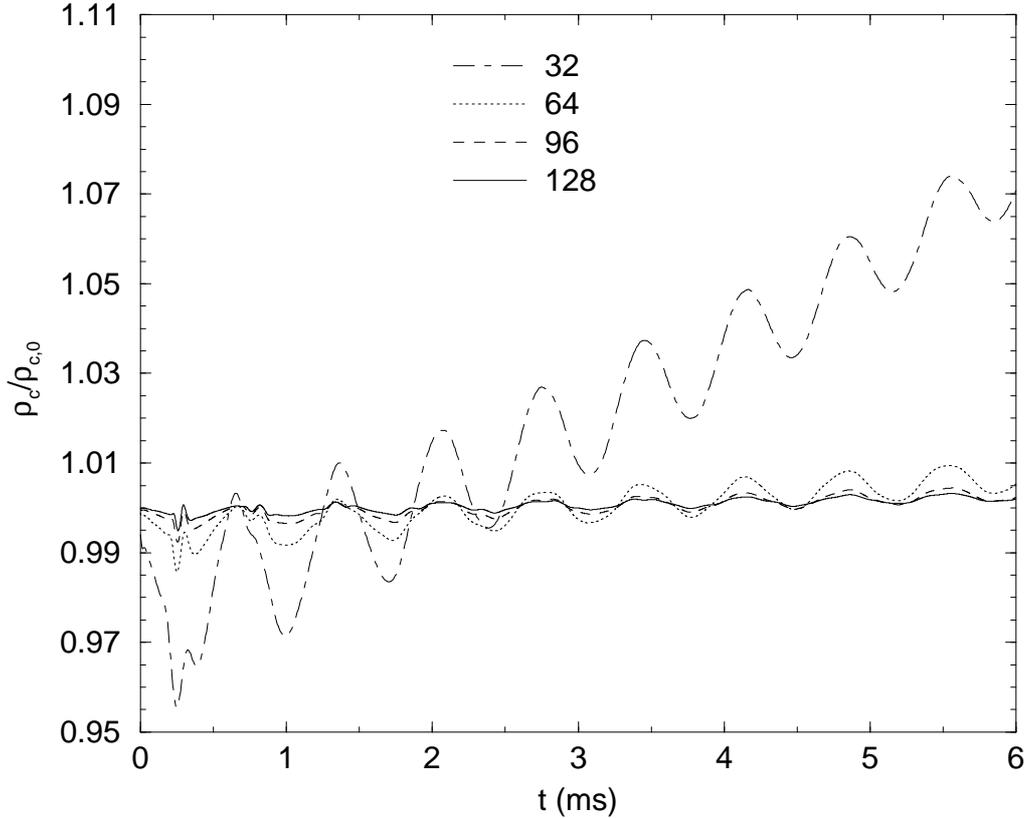}}}
      \caption{Central mass-density, normalized to the initial value, in a
   stable TOV star (\mbox{$ M = 1.4\, M_\odot $} and polytropic index $
   \Gamma = 2 $) evolution at different resolutions. PPM and Marquina
   were used for all runs.}  \label{figTOVdyn rho} \end{figure*}

\begin{figure}
\centering
\includegraphics[angle=0,width=6.5cm]{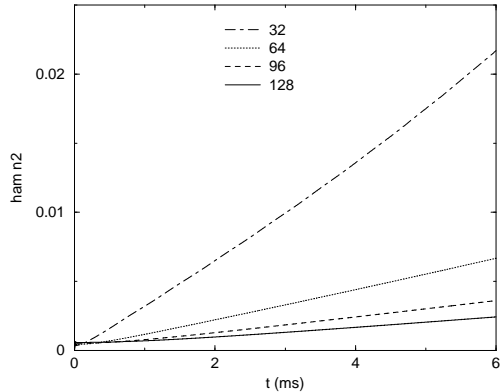}
\caption{The L2-norm of the Hamiltonian constraint for the same
evolutions as in Fig.~\ref{figTOVdyn rho}.}
\label{figTOVdyn ham}
\end{figure}

Next, we consider the evolution of stable relativistic polytropic
spherical (TOV) stars. As this is a static solution, no evolution
is expected. Yet as shown in Fig.~\ref{figTOVdyn rho}, both a small
periodic oscillation and a small secular increase of the central
density of the star are detected during the numerical evolution of the
equations. Both effects have, however, a single explanation. Since the
initial data contains also a small truncation error, this is
responsible for triggering radial oscillations which appear as
periodic variations in the central density. As the resolution is
increased, the truncation error is reduced and so is the amplitude of the
oscillation. The secular growth, on the other hand, is a purely
numerical problem, probably related to the violation of the constraint
equations. As for the oscillations, also the secular growth converges
to zero with increasing resolution. The convergence properties of the
code are also clearly shown in the growth of the Hamiltonian
constraint violation (Fig.~\ref{figTOVdyn ham}), where we can see that
almost second-order convergence is achieved. Note that the convergence
rate is not exactly second-order, because the reconstruction schemes
are only first-order accurate \citep{alcubierre et al} at local
extrema (i.e. the center and the surface of the star) thus increasing
the overall truncation error.

\begin{figure} 
\centering
\includegraphics[angle=0,width=6.6cm]{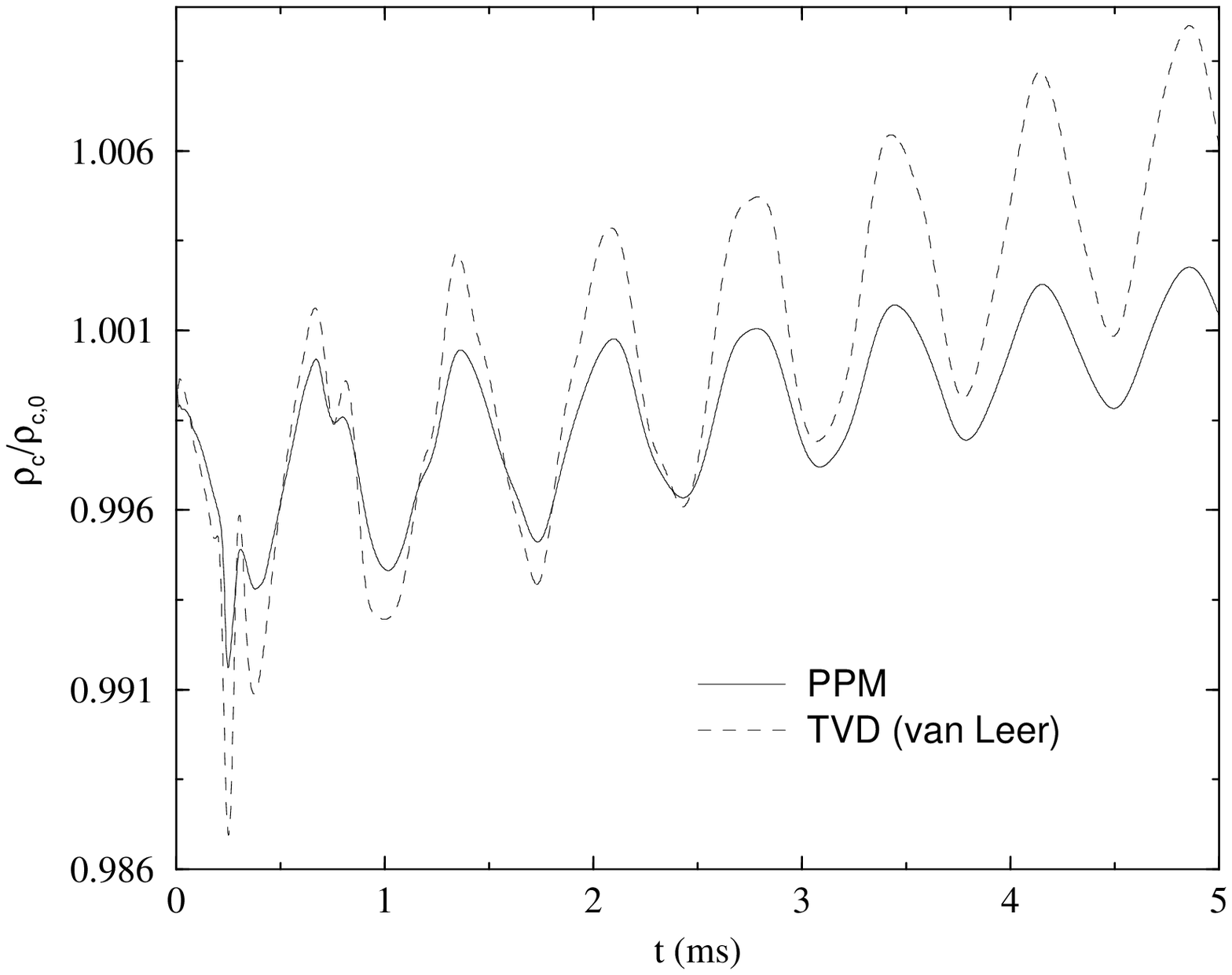}
\caption{Central mass-density, normalized to the initial value, in a stable
TOV star (\mbox{$ M = 1.4\, M_\odot $} and polytropic index $ \Gamma = 2
$) evolution with $ 64^3 $ grid points. Comparison between PPM and van
Leer reconstruction methods.}  \label{figCfPPM} 
\end{figure}

As anticipated, the PPM reconstruction scheme has shown to be more
accurate than the TVD ones. This is clear in Fig.~\ref{figCfPPM}, in
which the results obtained with the PPM method are compared to the
best of the TVD methods (i.e. the van Leer one) for a stable TOV run,
similar to the previous ones, using $ 64^3 $ grid points. Note that
the PPM reconstruction is more effective in reducing both the initial
truncation error (as shown by the smaller amplitudes in the
oscillations) and the secular error (as shown by the small growth
rate).

\begin{figure}
\centering
\includegraphics[angle=-90,width=6cm]{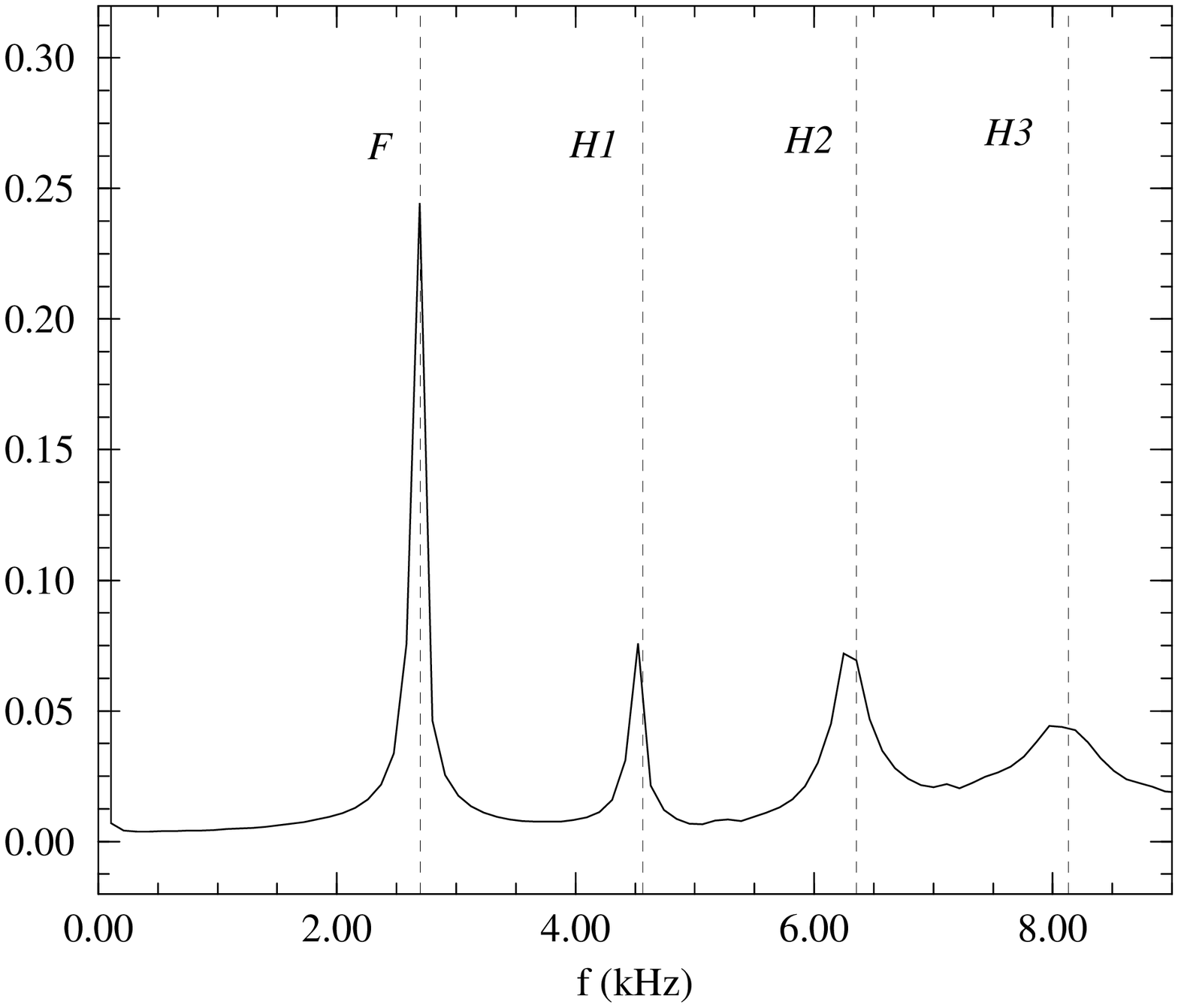}	
\caption{Fourier transform of the central mass-density evolution of an $
M = 1.4\, M_\odot $, $ \Gamma = 2 $ stable TOV star performed with $
128^3 $ grid points. The units of the vertical axis are arbitrary.}
\label{figFreq}
\end{figure}

In order to further investigate the accuracy of our implementation of
the hydrodynamics equations, we have suppressed the spacetime
evolution and solved just the hydrodynamics equations in the fixed
spacetime of the initial TOV solution. This approximation is referred
to as the ``Cowling approximation'' and is widely used in perturbative
studies of oscillating stars. In this case, in addition to the
confirmation of the convergence rate already checked in fully evolved
runs, we have also compared the frequency spectrum of the numerically
induced oscillation with the results obtained by an independent 2D
code \citep{2D code} and with perturbative analyses. In
Fig.~\ref{figFreq} we show a comparison between the two codes
reporting the power spectrum of the central density oscillations
computed with the {\tt Whisky} code and the corresponding frequencies
as obtained with perturbative techniques and with the 2D code. Clearly
the agreement is very good with an error below 1\% in the fundamental
frequency.  The fact that the frequencies computed with the code
coincide with the physical eigenfrequencies calculated through
perturbative analysis allows us to study with our code the physical
properties of linear normal-modes of oscillation even if such
oscillations are generated numerically.

The last test performed in the linear regime consisted of the
evolution of rapidly rotating stars, with angular velocity up to 95\%
of the allowed mass-shedding limit for uniformly rotating stars. The
initial data routines have been adapted from the {\tt RNS} code
\citep{RNS}. As in the previous tests the Hamiltonian constraint shows
a convergence rate of nearly second-order everywhere, except at the
surface and the center of the star. In analogy with the nonrotating
case, the truncation error triggers quasi-radial oscillations in the
star. Such pulsations converge to zero with increasing
resolution. Determining the frequency spectrum of fully relativistic
and rapidly rotating stars is an important scientific achievement,
allowing the investigation of a parameter space which is
astrophysically relevant but too difficult to treat with current
perturbative techniques.

Note that a number of small improvements on the boundary and gauge
conditions have allowed us to extend considerably the timescale of our
evolutions of stable rapidly rotating stars, which can now be evolved for
about 10 ms, that is for several rotational periods \citep{sterg priv}.

\begin{figure*}
\centering
\resizebox{\hsize}{!}{\rotatebox[]{0}{\includegraphics{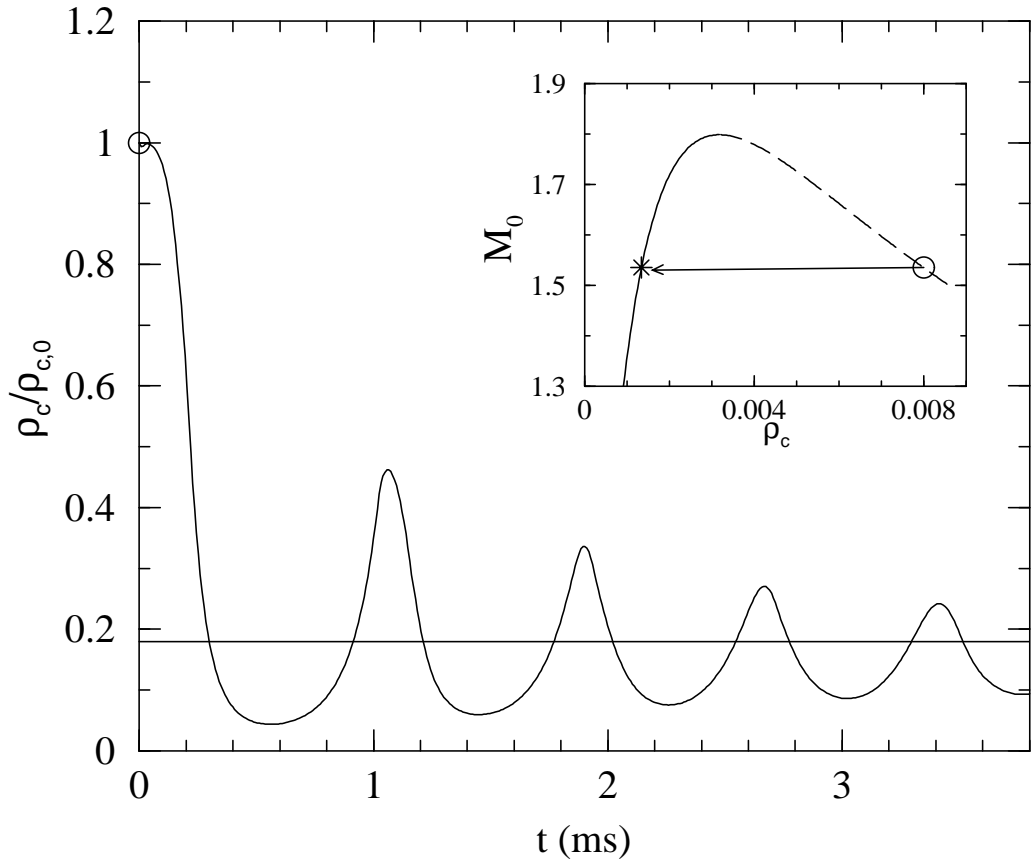}}}
\caption{Normalized central mass-density evolution of an $ M = 1.4\,
M_\odot $, $ \Gamma = 2 $ unstable TOV star performed with $ 96^3 $ grid
points.}
\label{figMigration}
\end{figure*}

We now consider tests of the nonlinear dynamics of isolated spherical
relativistic stars. To this purpose we have constructed TOV solutions
that are placed on the unstable branch of the equilibrium
configurations (see inset of Fig.~\ref{figMigration}). The truncation
error in the initial data for a TOV is sufficient to move the model to
a different configuration and in {\tt Whisky} this leads to a rapid
migration toward a stable configuration of equal rest-mass but smaller
central density. Such a violent expansion produces large amplitude
radial oscillations in the star that are either at constant amplitude
if the polytropic equation of state (\ref{poly}) is used, or are
damped through shock heating if the ideal fluid equation of state
(\ref{id fluid}) is used. This is summarized in
Fig.~\ref{figMigration}, which shows the time series of the normalized
central density for a TOV. We also show that the asymptotic central
density tends to a value corresponding to a rest-mass slightly smaller
than the initial one (straight line). This is the energy loss due to
the internal dissipation.

   \begin{figure}
   \centering
   \includegraphics[angle=0,width=6cm]{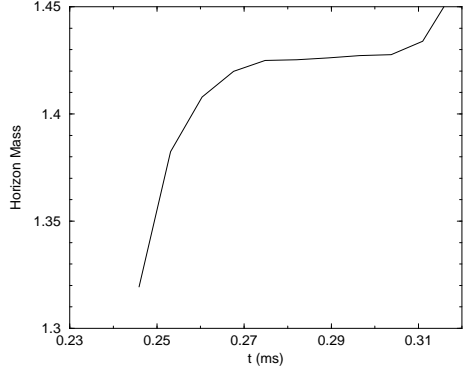}	
      \caption{Horizon mass evolution. The initial rest mass
   of the TOV star is $M = 1.44\, M_\odot$ and again $\Gamma = 2$.}
         \label{figCollapse}
   \end{figure}

An alternative solution for the unstable model is that of a gravitational
collapse. In this case, in fact, the initial star does not expand but
rather moves to increasingly larger density configurations, finally
forming a black hole. In order to study the gravitational collapse of the
unstable configuration, the introduction of a density perturbation in the
initial model is necessary. A very small one of the order of 1\% with
dependence $\cos(\pi r/2r_s)$, where $r$ is the coordinate distance from
the center and $r_s$ its value on the surface, is sufficient to overcome
the effects of the truncation error and induce the star to collapse. Note
that after adding the perturbation to the initial configuration, the
constraint equations are solved to provide initial data which are a
solution to the Einstein equations. As a summary of the results obtained,
we show in Fig.~\ref{figCollapse} the growth of the horizon mass, tracked
with an apparent horizon finder based on the fast-flow algorithm
\citep{horizon finder}. At $t = 0.246 $ ms a black hole forms and an
apparent horizon appears. As the remaining stellar material continues to
accrete onto the newly formed black hole, its horizon mass increases,
finally leveling off until $t = 0.306 $ ms. The subsequent growth of the
horizon mass is then only the result of the increasing error due to grid
stretching (the radial metric function develops a sharp peak which cannot
be resolved accurately enough).

\section{Conclusions}

We have illustrated the main features and the present status of our
new 3D general relativistic hydrodynamics code. Through a wide set of
numerical tests, we have shown both the accuracy and the convergence
of our implementation of the formulation of the general relativistic
hydrodynamics equations, coupled to a conformal transverse-traceless
formulation of the Einstein equations. We have also shown that our
code can accurately and stably evolve the linear and nonlinear
dynamics of isolated relativistic stars, both for pulsations of
spherical and rapidly rotating stars. The computed frequencies of
radial oscillations are compared with the corresponding frequencies
obtained with other numerical and perturbative techniques and the good
agreement among these values is one of the greatest present
achievements of our code.

We have also investigated the nonlinear dynamics of stellar models
that are unstable to the fundamental radial mode of pulsation,
decaying either to a black hole (a collapse which we can follow up to
and past the formation of the event horizon) or in a migration to a
configuration on the stable branch of equilibrium configurations.

These encouraging results are important premises for the application
of the {\tt Whisky} code to more physical scenarios. Furthermore we
are presently working on incorporating in {\tt Whisky} the use of a
fixed mesh refinement. This is an important improvement that will
increase the numerical resolution where needed and move the position
of the outer boundaries further out in the wave zone, where
information on the gravitational wave content of the spacetime can be
reliably extracted.

\begin{acknowledgements}
It is a pleasure to thank J. Frieben, J. A. Font, {Jos\'e
M$^{\underline{\mbox{a}}}$. Ib\'a\~{n}ez}, F. L\"offler, E. Seidel and
N. Stergioulas, who have participated to the development and testing of
the code. Special thanks also go to Roberto Capuzzo Dolcetta for
promoting the first of this series of conferences. Financial support for
this work has been provided by the MIUR and EU Network Programme
(Research Training Network contract HPRN-CT-2001-01172). The computations
were performed on the Beowulf Cluster for numerical relativity {\it
``Albert100''}, at the University of Parma.

\end{acknowledgements}

\bibliographystyle{aa}

\end{document}